\documentclass[aps, preprint, showpacs,groupedaddress]{revtex4-1}
\usepackage{dcolumn}
\usepackage{graphicx}
\usepackage{bm, amsmath, amssymb, textcomp}
\usepackage{hyperref}

\begin{document}

\title{Quasiparticle band structures and thermoelectric transport properties of p-type SnSe}

\author{Guangsha Shi}
\author{Emmanouil Kioupakis}
\affiliation{Department of Materials Science and Engineering, University of Michigan, Ann Arbor, Michigan 48109, USA}

\date{\today}

\begin{abstract}

We used density functional and many-body perturbation theory to calculate the quasiparticle band structures and electronic transport parameters of p-type SnSe
both for the low-temperature \emph{Pnma} and high-temperature \emph{Cmcm} phases.
The \emph{Pnma} phase has an indirect band gap of 0.829 eV while the \emph{Cmcm} has a direct band gap of 0.464 eV.
Both phases exhibit multiple local band extrema within an energy range comparable to the thermal energy of carriers from the global extrema.
We calculated the electronic transport coefficients
for single-crystal and polycrystalline materials to understand previous experimental measurements.
We also discuss the dependence of the transport coefficients on doping concentration and temperature
to identify doping conditions for optimal thermoelectric performance.
\pacs{71.20.Nr,72.20.Pa,84.60.Rb}
\end{abstract}

\maketitle

\section{Introduction}
\begin{figure}
\includegraphics[width=\columnwidth]{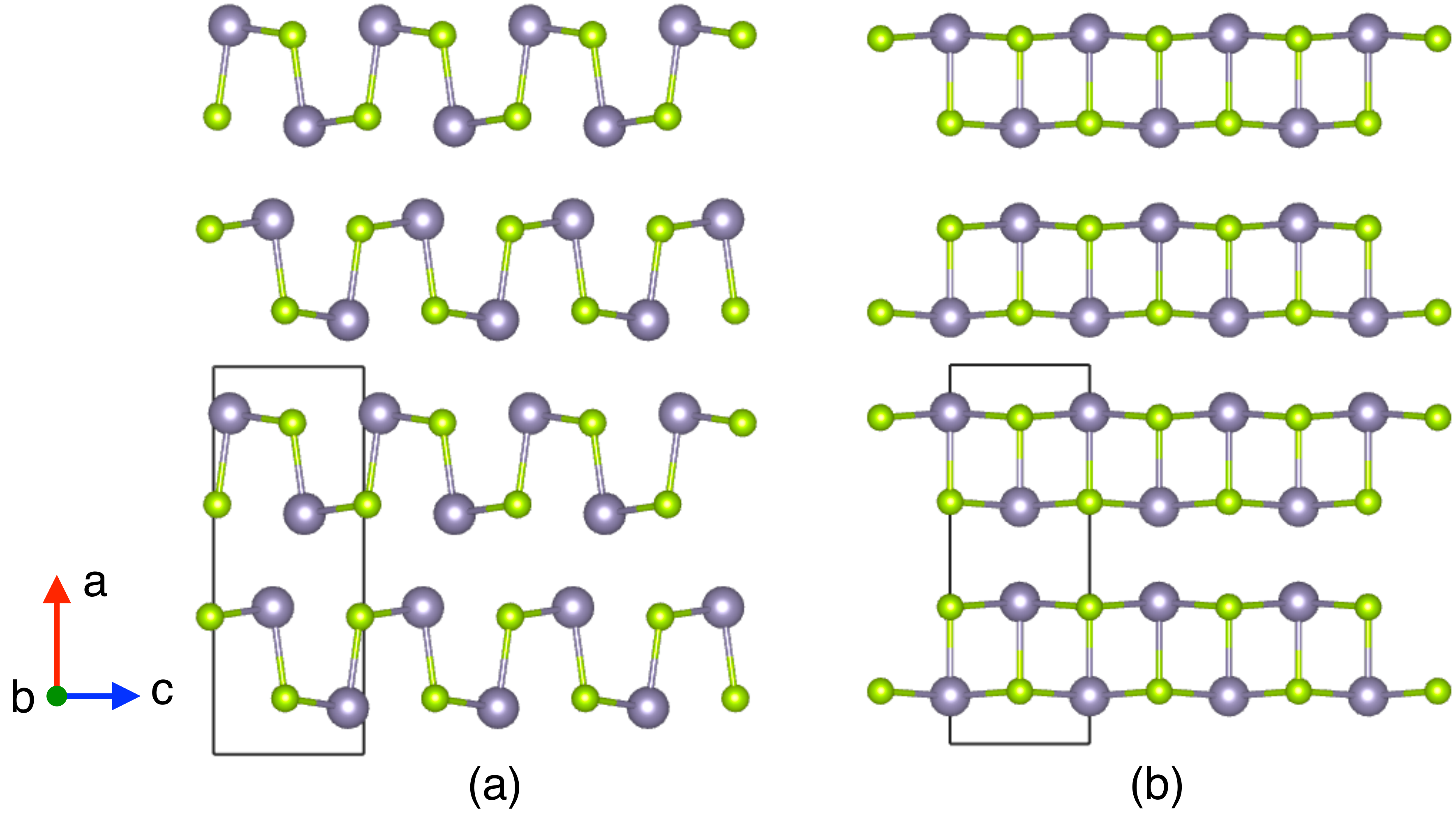}
\caption{
\label{fig:SnSe_structure}
Crystal structures of (a) the low-temperature (\emph{Pnma}) phase and (b) the high-temperature (\emph{Cmcm}) phase of SnSe.
}
\end{figure}

Thermoelectric materials enable the direct conversion of heat to electricity and can be used to recover usable energy from waste heat.
The efficiency of thermoelectric energy conversion is determined by the dimensionless figure of merit of the material, $ZT=S^2\sigma T/(\kappa_{\text{L}}+\kappa_{\text{el}})$,
where $S$ is the Seebeck coefficient, $\sigma$ is the electrical conductivity, $T$ is the absolute temperature,
and $\kappa_{\text{L}}$ and $\kappa_{\text{el}}$ are the lattice and electronic contributions to the thermal conductivity.
High $ZT$ values occur in materials with high electrical conductivity, high Seebeck coefficient, and low thermal conductivity,
such as p-type IV-VI compounds (i.e., PbSe, PbTe, and their alloys) with reported $ZT$ values as high as 1.8.\cite{Heremans25072008,C0EE00456A,pei2011convergenceof}
SnSe is another IV-VI compound that has attracted little attention as a thermoelectric material.
Recently, Zhao \emph{et al.} reported figure-of-merit values as high as 2.6
in single-crystal samples of unintentionally doped p-type SnSe.\cite{zhao2014ultralowthermal}
SnSe undergoes a phase transition at 813 K from the low-temperature
phase (\emph{Pnma} space group) to the high-temperature \emph{Cmcm} phase (Fig.~\ref{fig:SnSe_structure}).
The highest $ZT$ values were found near and above this phase-transition temperature of 813 K.

In this work we present the quasiparticle band structures and thermoelectric transport coefficients of
both the low-temperature (\emph{Pnma}) and high-temperature (\emph{Cmcm}) phases of SnSe.
In Section~\ref{sec:methodology} we discuss our first-principles methodology for the calculation of the band structures and transport coefficients.
In Section~\ref{sec:results_and_discussion} we present and discuss our findings.
In Section~\ref{sec:band_structure}
we report values for the band gaps, band-extrema locations, and carrier effective masses for both phases.
We found multiple local band extrema that lie close in energy to the the band edges and
need to be considered when calculating the thermoelectric transport properties.
In Section~\ref{sec:transport_experiment}
we calculate the Seebeck coefficient and electrical conductivity of both SnSe phases and compare to 
recent experimental measurements.
In Section~\ref{sec:n_T_dep} we report
the carrier-density and temperature dependence of the electronic transport coefficients.
We predict that SnSe shows good thermoelectric performance at high temperature
when doped in the 10$^{19}$--10$^{20}$ cm$^{-3}$ range.

\section{Methodology}\label{sec:methodology}
We studied the electronic properties of SnSe using first-principles calculations based on density functional and many-body perturbation theory.
We calculated the ground-state charge density and electronic wave functions
using the generalized gradient approximation\cite{PhysRevLett.45.566,PhysRevLett.77.3865} for the exchange-correlation potential.
We used the plane-wave pseudopotential method\cite{0022-3719-12-21-009}
with norm-conserving pseudopotentials\cite{TroullierMartins91} and a plane-wave cutoff of 200 Ry as implemented in the Quantum-ESPRESSO code\cite{0953-8984-21-39-395502}.
The electrons from the outermost valence shell (5\emph{s} and 5\emph{p}) as well as those from the semicore atomic shell (4\emph{s}, 4\emph{p}, and 4\emph{d})
are included for Sn. 
The crystal structures of the two phases are shown in Fig.~\ref{fig:SnSe_structure} along the \emph{b} axis.
We used the experimentally measured values for the lattice vectors and atomic positions of SnSe
at 300K for the low-temperature \emph{Pnma} phase (\emph{a} = 11.50 \AA, \emph{b} = 4.45 \AA, \emph{c} = 4.153 \AA)
and at 813 K for the high-temperature \emph{Cmcm} (\emph{a} = 4.31 \AA, \emph{b} = 6.24 \AA, \emph{c} = 4.31 \AA) phase.\cite{1998ZK....213..343A}
The Brillouin zone was sampled using a Monkhorst-Pack grid\cite{PhysRevB.13.5188}
of 6$\times$6$\times$2 for the low-temperature phase and 6$\times$6$\times$4 for the high-temperature phase.

We calculated the quasiparticle band structure of SnSe using the one-shot GW method\cite{PhysRevB.34.5390} and the BerkeleyGW code\cite{Deslippe20121269}. The static dielectric function was calculated with a 20 Ry plane-wave cutoff and extended to finite frequency using the generalized plasmon-pole model of Hybertsen and Louie\cite{PhysRevB.34.5390}. The Coulomb-hole self-energy term was calculated using a sum over unoccupied bands up to 16 Ry above the valence band maximum using the static-remainder approach\cite{PhysRevB.87.165124}. Corrections due to spin-orbit coupling interaction\cite{PhysRevB.34.2920} were calculated in a non-self-consistent way using plane waves up to a cut-off energy of 50 Ry. We used the maximally localized Wannier function formalism\cite{RevModPhys.84.1419,Mostofi2008685} to interpolate the quasiparticle energies and spin-orbit coupling matrix elements to arbitrary points in the first Brillouin zone as in Ref.~\onlinecite{PhysRevB.82.245203}. Subsequently, we interpolated the quasiparticle energies 
on fine meshes in the first Brillouin zone (120$\times$120$\times$60 for \emph{Pnma}-SnSe and 120$\times$120$\times$80 for \emph{Cmcm}-SnSe)
to determine the thermoelectric transport coefficients of p-type SnSe in the constant-relaxation-time approximation.\cite{PhysRevB.68.125210,PhysRevB.81.155211}



\begin{figure*}
\includegraphics[width=\textwidth]{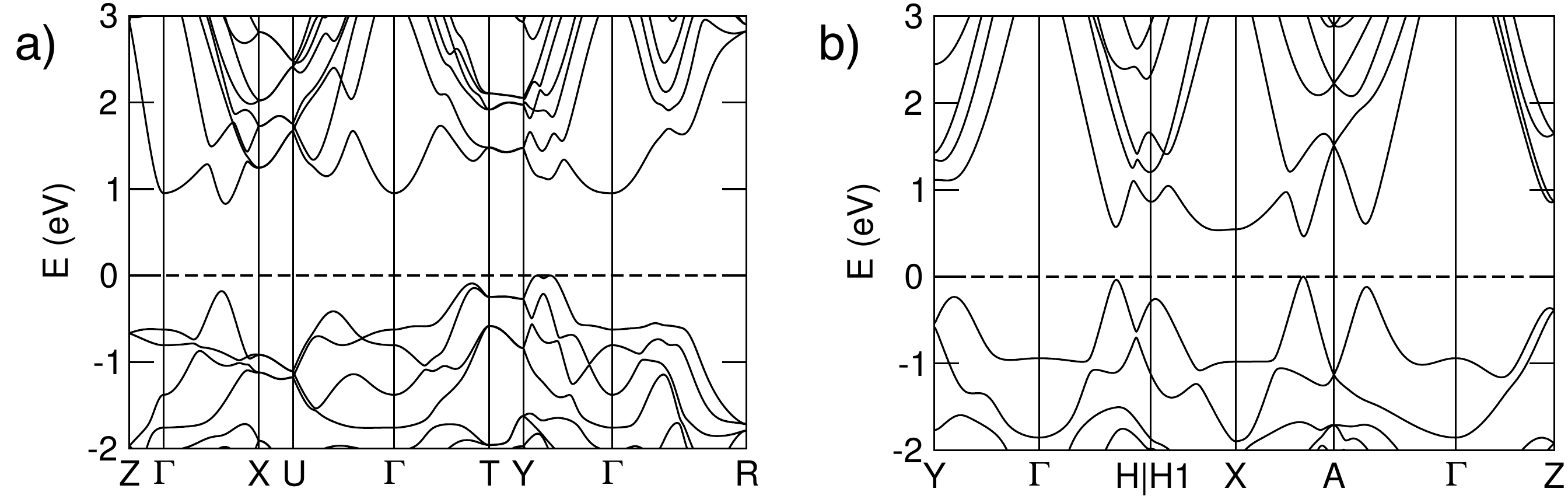}
\caption{
\label{fig:SnSe_band}
Quasiparticle band structures of (a) the low-temperature (\emph{Pnma}) phase and (b) the high-temperature (\emph{Cmcm}) phase of SnSe. \emph{Pnma}-SnSe has an indirect band gap of 0.829 eV, while the band gap of \emph{Cmcm}-SnSe is direct with a magnitude of 0.464 eV. Both phases exhibit multiple local band extrema that lie close in energy to the global extrema. }
\end{figure*}

\section{Results and discussion}\label{sec:results_and_discussion}

\subsection{Band structure}\label{sec:band_structure}
The calculated band structures of \emph{Pnma}-SnSe and \emph{Cmcm}-SnSe, including quasiparticle and spin-orbit coupling corrections, are shown in Figs.~\ref{fig:SnSe_band}(a) and~\ref{fig:SnSe_band}(b). The band gap of \emph{Pnma}-SnSe is found to be indirect with a magnitude of 0.829 eV, while \emph{Cmcm}-SnSe has a direct band gap with a magnitude of 0.464 eV. 
The Brillouin-zone positions, energies, and multiplicities of the band extrema of \emph{Pnma}-SnSe and \emph{Cmcm}-SnSe
are summarized in Table~\ref{tab:eff_mass_pnma}. 
All energies are referenced to the valence-band maximum (VBM) of each phase.
In \emph{Pnma}-SnSe, the position of the VBM is at (0.00, 0.35, 0.00) along the $\Gamma$--Y direction of the first Brillouin zone.
There is also a local valence-band maximum at (0.00, 0.42, 0.00) that lies within 1 meV lower in energy than the VBM.
The conduction-band minimum (CBM) is located at (0.33, 0.00, 0.00) along the $\Gamma$--X direction.
The calculated band gap (0.829 eV) is in good agreement with optical-absorption measurements (0.86 eV\cite{zhao2014ultralowthermal} and 0.898 eV\cite{PhysRevB.41.5227}).
The smallest direct transition energy was found to be 1.03 eV and occurs at (0.32, 0.00, 0.00), which is close to the CBM location.
In \emph{Cmcm}-SnSe, both the VBM and CBM are located at (0.34, 0.50, 0.00) along the X--A direction.
In addition, the band structure exhibits a local valence-band maximum, VBM1 at (0.00, 0.20, 0.39)
along the $\Gamma$--H direction, located just 31 meV lower in energy than the global VBM.
A local conduction-band minimum, CBM1 is found at (0.00, 0.54, 0.08) along the X--H1 direction, located 70 meV higher in energy than the global CBM.

\begin{figure*}
\includegraphics[width=\textwidth]{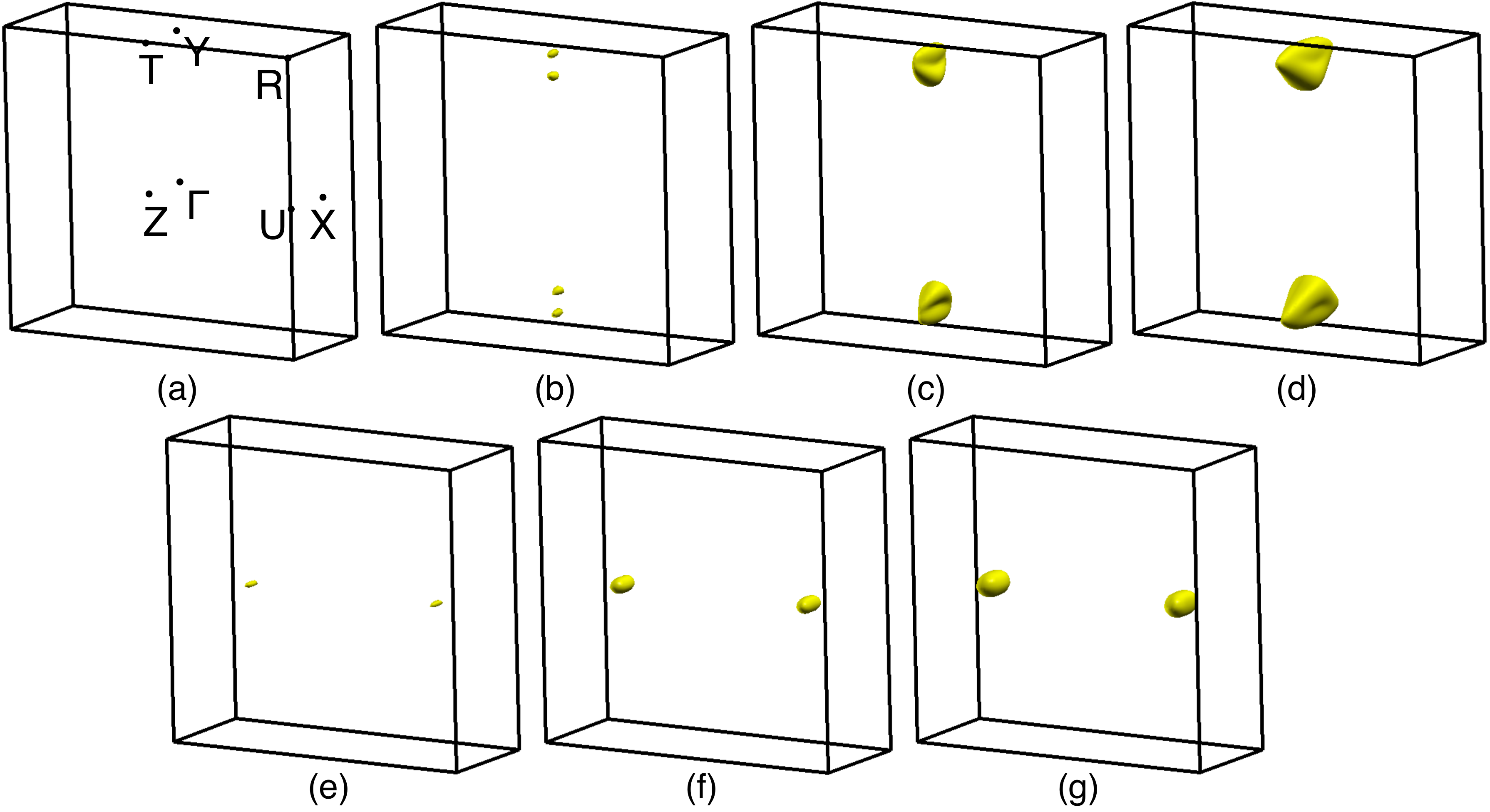}
\caption{
\label{fig:SnSe_alpha_BZ}
(a) The first Brillouin zone of the low-temperature (\emph{Pnma}) phase of SnSe.  (b$\sim$d) Constant-energy surfaces of the highest valence band with an energy of (b) 10 meV, (c) 50 meV, (d) 100 meV lower than the VBM energy. (e$\sim$g)  Constant-energy surfaces of the lowest conduction band with an energy of (e) 10 meV, (f) 50 meV, (g) 100 meV higher than the CBM energy.
}
\end{figure*}

\begin{figure*}
\includegraphics[width=\textwidth]{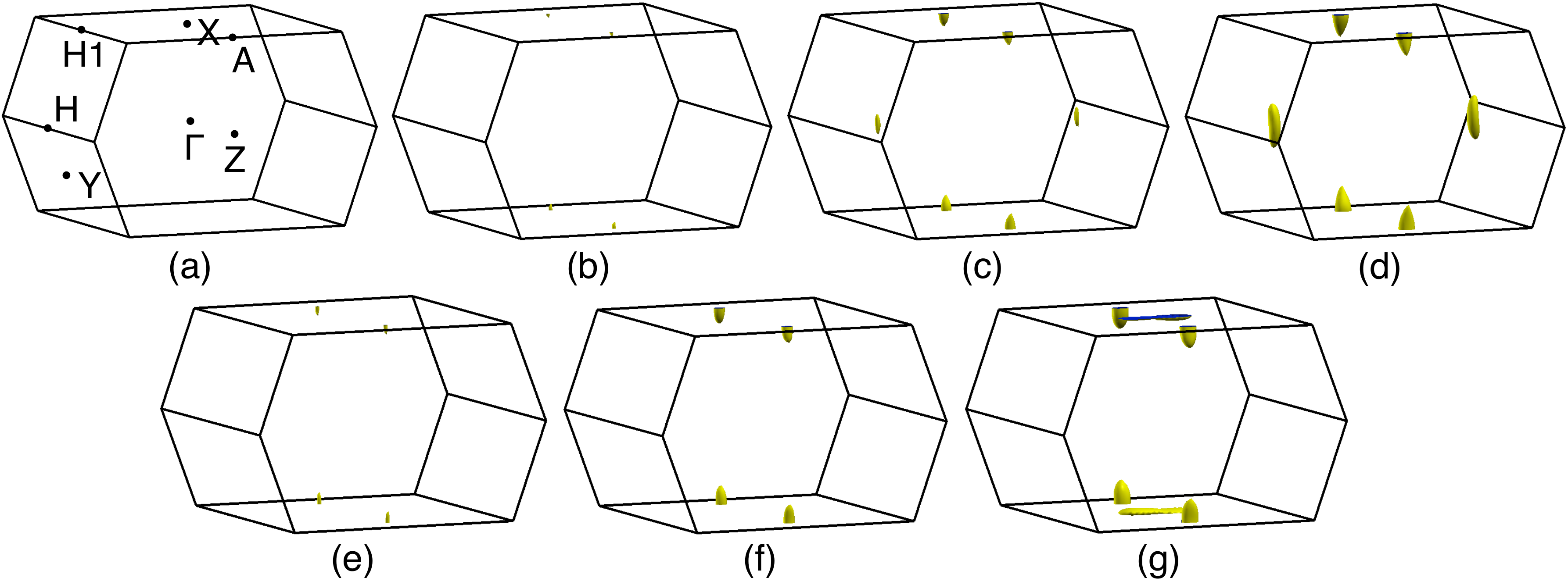}
\caption{
\label{fig:SnSe_beta_BZ}
(a) The first Brillouin zone of the high-temperature (\emph{Cmcm}) phase of SnSe.  (b$\sim$d) Constant-energy surfaces of the highest valence band with an energy of (b) 10 meV, (c) 50 meV, (d) 100 meV lower than the VBM energy. (e$\sim$g)  Constant-energy surfaces of the lowest conduction band with an energy of (e) 10 meV, (f) 50 meV, (g) 100 meV higher than the CBM energy.
}
\end{figure*}

Figures \ref{fig:SnSe_alpha_BZ} and \ref{fig:SnSe_beta_BZ} show a set of constant-energy surfaces of the valence and conduction bands for both 
SnSe phases plotted within the first Brillouin zone. 
The plots demonstrate the multiple local extrema for both phases. The local extrema reside within an energy range of $k_{\text{B}}T$ ($k_{\text{B}}$ is the Boltzmann constant)
from the band edges for temperatures near the phase-transition temperature (813 K) around which the highest $ZT$ values have been reported. Therefore, all local extrema need to be taken into account when analyzing the transport properties of n-type and p-type SnSe.

The effective-mass parameters are also reported for both phases in Table~\ref{tab:eff_mass_pnma}. The effective mass at each extremum is highly anisotropic.
With the exception of the CBM1 local minimum of the \emph{Cmcm} phase, the effective mass has a larger value along the $a$ axis, perpendicular to the atomic layers,
than either of the in-plane directions $b$ and $c$. This is due to the two-dimensional nature of the material, which favors electron transport within the atomic layers than
perpendicular to them.

\begin{table}[b]
\caption{
\label{tab:eff_mass_pnma}
Calculated values of the positions and energies of the conduction and valence band extrema,
and effective masses along the \emph{a}, \emph{b}, and \emph{c} axes for the low-temperature (\emph{Pnma}) and high-temperatute (\emph{Cmcm}) phase of SnSe.
The positions (k1, k2, k3) are in crystal coordinates. The energies are referenced to the energy of the VBM for each phase.
}
\begin{ruledtabular}
\begin{tabular}{ccccccc}
&Multiplicity&(k1, k2, k3)&E (eV)&$m_{a}^{*}$&$m_{b}^{*}$&$m_{c}^{*}$\\
\hline
\emph{Pnma} & & & & & & \\
\hline
VBM & 2 & (0.00, 0.35, 0.00) & 0.000 & 0.74 & 0.31 & 0.16\\
VBM1 & 2 & (0.00, 0.42, 0.00) & 0.000 & 0.90 & 0.12 & 0.15\\
CBM & 2 & (0.33, 0.00, 0.00) & 0.829 & 2.40 & 0.11 & 0.15\\
\hline
\emph{Cmcm} & & & & & & \\
\hline
VBM & 2 & (0.34, 0.50, 0.00) & 0.000 & 0.34 & 0.04 & 0.09\\
VBM1 & 2 & (0.00, 0.20, 0.39) & -0.031 & 0.77 & 0.12 & 0.05\\
CBM & 2 & (0.34, 0.50, 0.00) & 0.464 & 3.07 & 0.04 & 0.10\\
CBM1 & 2 & (0.00, 0.54, 0.08) & 0.534 & 0.06 & 0.82 & 1.52\\
\end{tabular}
\end{ruledtabular}
\end{table}


\subsection{Transport coefficients}\label{sec:transport_experiment}

\begin{figure*}
\includegraphics[width=\textwidth]{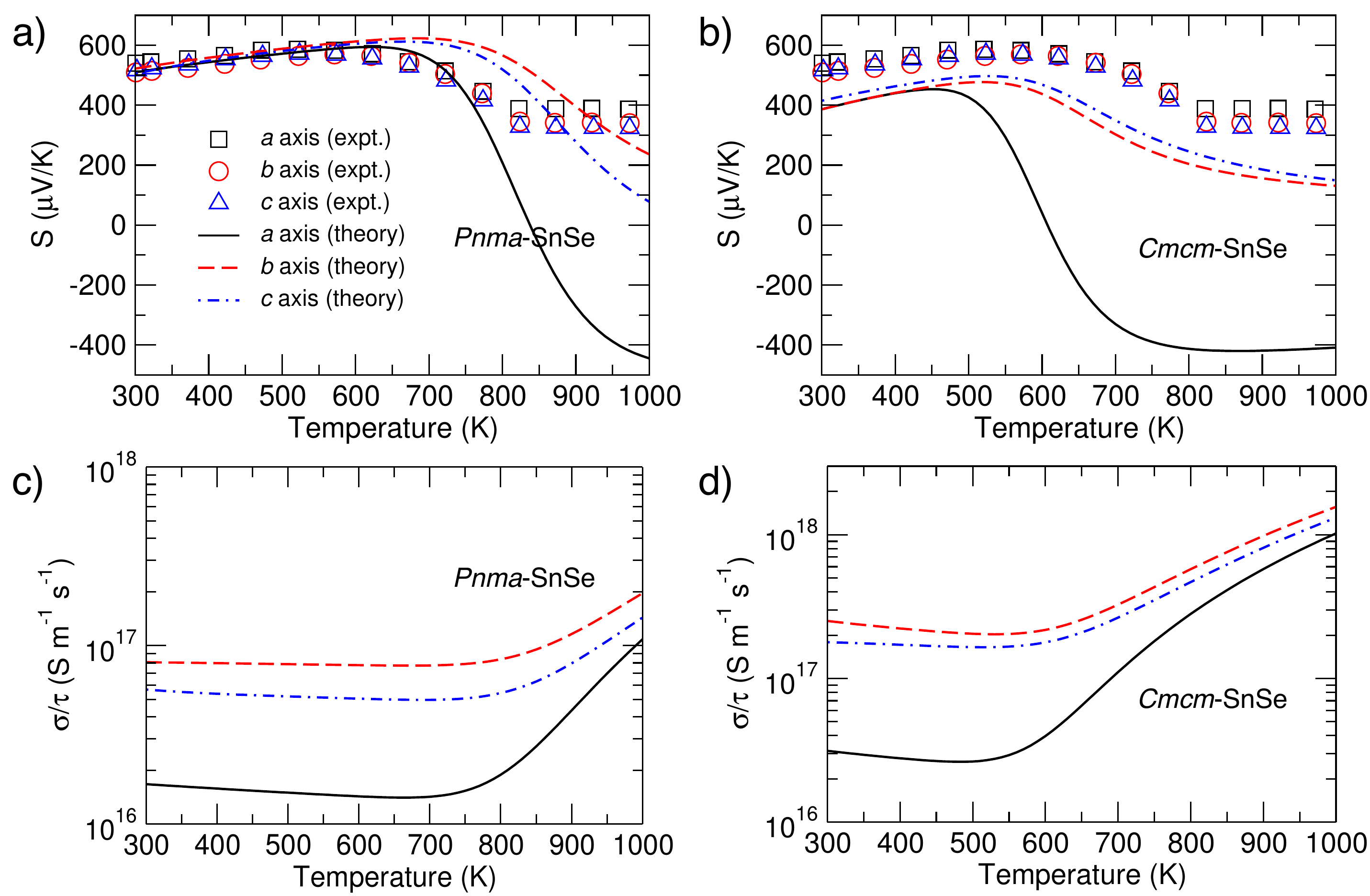}
\caption{
\label{fig:SnSe_SandSigma_6E17}
Calculated values (lines) for the Seebeck coefficient $S$ (a, b) and the electrical conductivity divided by the constant scattering time $\sigma/\tau$ (c, d)
of the  low-temperature (\emph{Pnma}) phase and high-temperature (\emph{Cmcm}) phase of SnSe for a doping concentration (net free-carrier concentration)
of 6.0$\times$$10^{17}$cm$^{-3}$,
which matches the experimental Hall coefficient measurements at 300 K in Ref.~\onlinecite{zhao2014ultralowthermal}.
}
\end{figure*}

Figures~\ref{fig:SnSe_SandSigma_6E17}(a) and~\ref{fig:SnSe_SandSigma_6E17}(b) show the calculated Seebeck coefficients for single-crystal 
\emph{Pnma}-SnSe and \emph{Cmcm}-SnSe
as a function of crystal direction and temperature.
The calculated data are also compared to the experimental values reported by Zhao \emph{et al.} for single-crystal samples\cite{zhao2014ultralowthermal}.
For these calculations we assumed a doping concentration (i.e., net free-carrier concentration) of 6$\times$10$^{17}$ cm$^{-3}$,
which agrees with the experimental Hall measurements at 300 K.\cite{zhao2014ultralowthermal}
Although the two phases of SnSe are stable in different temperature regimes,
we present theoretical results for the transport coefficients for both phases at all temperatures for completeness.
In the 300--600 K temperature range the calculated Seebeck coefficients for the \emph{Pnma} phase increase with temperature and 
are in good agreement with the experimental data [Fig.~\ref{fig:SnSe_SandSigma_6E17}(a)].
For temperatures in the range of 600--813 K electrons are thermally excited across the gap and induce bipolar transport, which reduces the Seebeck coefficient.
Our theoretical Seebeck-coefficient data for the \emph{Pnma} phase along the $b$ and $c$ axes are larger than experiment in this temperature regime
because our theory predicts that the onset of bipolar transport
occurs at higher temperatures than experiment [Fig.~\ref{fig:SnSe_SandSigma_6E17}(a)].
This is because we have not included the effect of temperature on the calculated band structure, which in general decreases
the band gap with increasing temperature and reduces the temperature onset of bipolar transport.
The theoretical Seebeck coefficient for the \emph{Pnma} phase along the $a$ axis decreases rapidly as a function of temperature above 600 K and
eventually changes sign around 840 K, which is above the phase-transition temperature of 813 K. 
We attribute the rapid decrease and sign reversal of the Seebeck coefficient along the $a$ axis in this temperature range
to the increasing negative contribution of thermally excited electrons to the Seebeck coefficient under bipolar-transport conditions.
The calculated Seebeck coefficient data for the \emph{Cmcm} phase in Fig.~\ref{fig:SnSe_SandSigma_6E17}(b) show that bipolar transport sets in 
at lower temperatures than in \emph{Pnma}-SnSe because \emph{Cmcm}-SnSe has a lower band gap.
Moreover, our calculations predict that for the experimental doping level of 6$\times$10$^{17}$ cm$^{-3}$ the sign of the Seebeck coefficient
along the $a$ direction becomes negative for temperatures above 600 K, and in particular in the 813--1000 K temperature range where the \emph{Cmcm} phase is
stable. The experimental Seebeck coefficient values for the \emph{Cmcm} phase along the $b$ and $c$ directions are larger than our calculated results,
while no sign reversal of the Seebeck coefficient along the $a$ direction is observed experimentally.
We discuss the potential origins of this discrepancy later in this Section.

\begin{figure}
\includegraphics[width=\columnwidth]{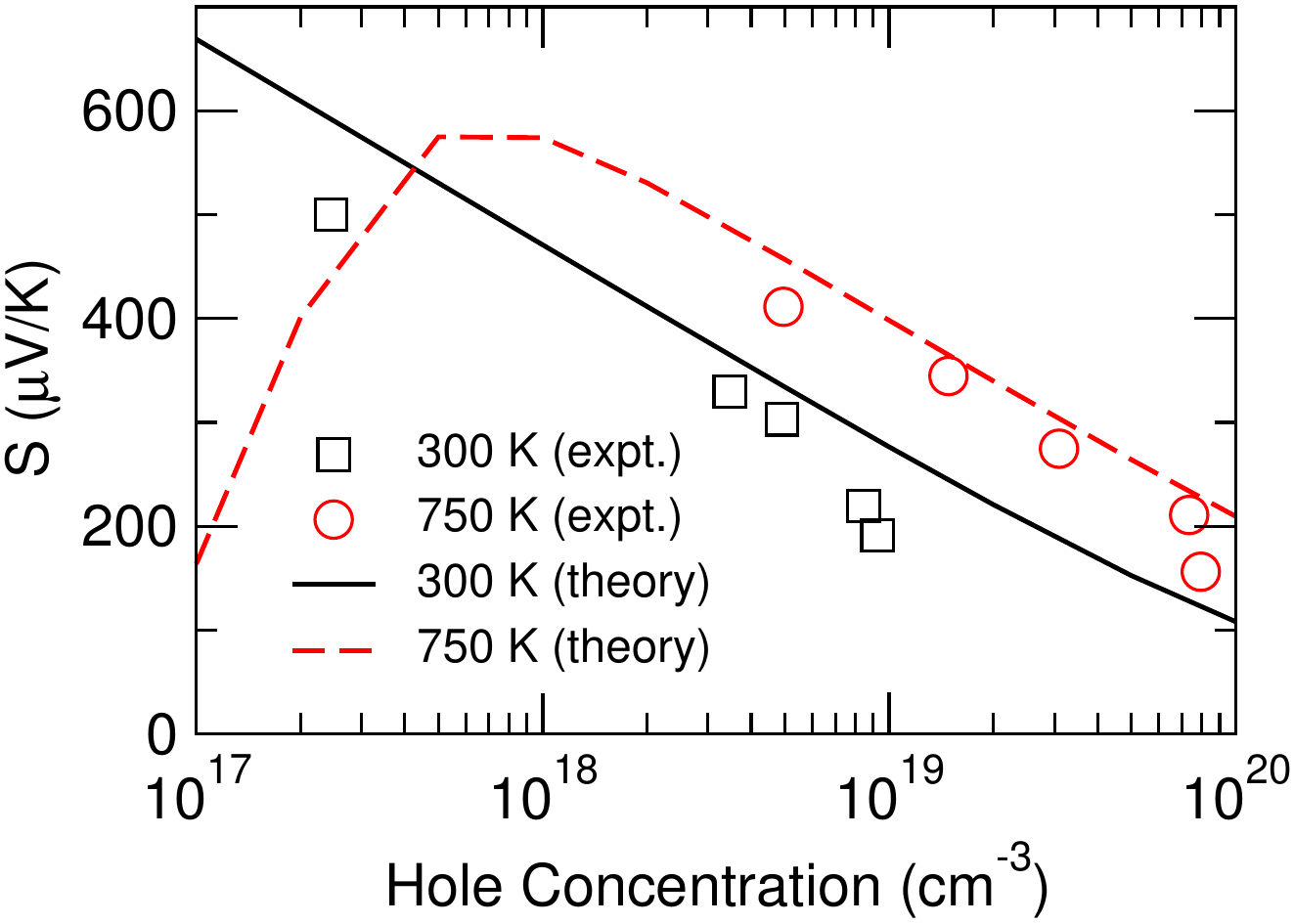}
\caption{
\label{fig:SnSe_poly_seebeck_N}
Directionally averaged Seebeck coefficient of \emph{Pnma}-SnSe as a function of net free-carrier concentration and temperature. The calculated data are in very good agreement with 
the experimental results for polycrystalline SnSe reported in Ref.~\onlinecite{C4TA01643B}.
}
\end{figure}

We also calculated the Seebeck coefficients of polycrystalline \emph{Pnma}-SnSe as a function of carrier density and temperature and compared to experiment.
The Seebeck coefficients for polycrystalline \emph{Pnma}-SnSe are evaluated by calculating the directional average  
along the \emph{a}, \emph{b}, and \emph{c} axes weighted by the electrical conductivity according to
$S_\text{avg} = (S_a\sigma_a+S_b\sigma_b+S_c\sigma_c)/(\sigma_a+\sigma_b+\sigma_c)$.
The calculated Seebeck data are plotted as a function of hole concentration for two temperatures (300 K and 750 K) in Fig.~\ref{fig:SnSe_poly_seebeck_N}.
The calculated coefficients are found to be in very good agreement with the recent 
experimental work by Chen \emph{et al.}
for p-type polycrystalline SnSe doped with Ag.\cite{C4TA01643B}

Figures \ref{fig:SnSe_SandSigma_6E17}(c) and \ref{fig:SnSe_SandSigma_6E17}(d) show the calculated electrical conductivity divided by the scattering time
for the \emph{Pnma} and \emph{Cmcm} phases of single-crystal SnSe
as a function of temperature and crystal direction.
The net free-carrier concentration is taken to be 6$\times$10$^{17}$ cm$^{-3}$,
equal to the experimental Hall measurements in the work of Zhao \emph{et al.} at 300 K.\cite{zhao2014ultralowthermal}
The electrical conductivity is similar along the \emph{b} and \emph{c} axes but much smaller along the \emph{a} axis.
This trend is in good agreement with experiment\cite{zhao2014ultralowthermal} and stems from the anisotropic two-dimensional nature of the material.
In the 300--700 K temperature range the calculated data for the \emph{Pnma} phase do not depend strongly on temperature,
indicating that the free-carrier concentrations do not change 
substantially in this temperature regime. For temperatures above 700 K bipolar transport sets in and both
the number of thermally excited carriers and the electrical conductivity increase exponentially with temperature.
The electrical conductivity for the \emph{Cmcm} phase has a similar behavior to the \emph{Pnma} phase,
with the difference that due to the smaller band gap of \emph{Cmcm}-SnSe bipolar transport starts to occur at lower temperatures around 500 K.
The experimental data for the \emph{Pnma} phase show a temperature dependence qualitatively similar to our calculations.
The electrical conductivity decreases weakly with temperature in the 300--550 K range,
while for temperatures above 550 K it increases exponentially as carriers are thermally excited across the gap. 
Remarkably, the experimental conductivity data for the \emph{Cmcm} phase show a weak decrease with temperature above
the phase-transition temperature of 813 K, in sharp contrast with the exponential increase observed for temperatures
above 500 K in our calculations.

There are several discrepancies between our theoretical calculations and the experimental data for the Seebeck coefficient and
electrical conductivity of SnSe above the phase-transition temperature of 813 K.
Although the calculated Seebeck coefficients are in very good agreement with the low-temperature \emph{Pnma} phase experimental data,
our calculations do not reproduce the approximately constant Seebeck coefficients reported experiemntally for the \emph{Cmcm} phase above 813 K.\cite{zhao2014ultralowthermal}
Moreover, the origin of the experimentally observed behavior of the electrical conductivity for the \emph{Cmcm} phase is not clear.
We expect the number of thermally excited carriers for the \emph{Cmcm} phase
to increase exponentially with temperature above 500 K due to bipolar transport.
It is not obvious what is the cause of the discrepancy between theory and experiment at high temperatures.
On the one hand, the disagreement could be attributed to limitations of our computational method. Our calculations assume the relaxation time to 
be constant, isotropic, and the same for both electrons and holes.
The relaxation time may be very different between electrons and holes, and it may also vary with direction, energy, and temperature.
Moreover, we have not explicitly considered temperature effects on the energy eigenvalues.
However, the good agreement
between our calculated data and experiment for single-crystal and polycrystalline \emph{Pnma}-SnSe
suggests that our calculated band structures are accurate and the constant relaxation time is a valid approximation.
On the other hand, the increasing temperature and the phase transition may affect the nature or concentration of defects and dopant impurities in the material
and thus the net free-carrier concentration.
Indeed, it is hard to identify a different reason why the conductivity does not continue to increase
exponentially with temperature above the transition to the lower-band-gap \emph{Cmcm} phase as reported experimentally.\cite{zhao2014ultralowthermal}
Further evidence for this point is provided by our data for the upper limit of $ZT$ discussed later. 

\begin{figure}
\includegraphics[width=\columnwidth]{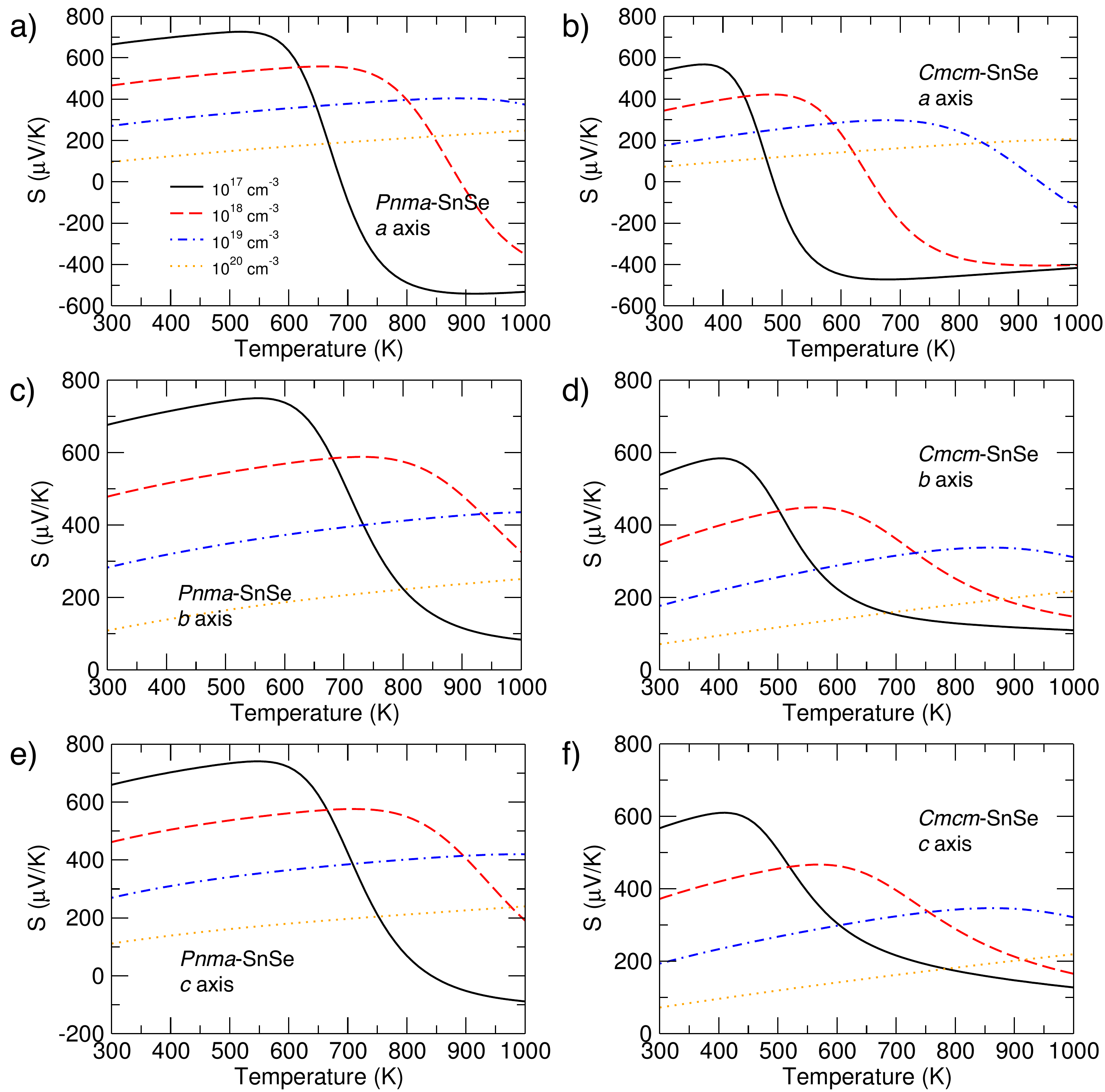}
\caption{
\label{fig:S_ptype}
Seebeck coefficient as a function of temperature and carrier concentration for the low-temperature (\emph{Pnma}) (a, c, e) and high-temperature (\emph{Cmcm}) phases (b, d, f) of SnSe along the \emph{a}, \emph{b}, and \emph{c} axes of the crystal structure.
}
\end{figure}

\subsection{Carrier-density and temperature dependence of transport coefficients}\label{sec:n_T_dep}

To further explore the optimal temperature and carrier concentration for the best thermoelectric performance,
we calculated the thermoelectric transport properties of both \emph{Cmcm} and \emph{Pnma} phases
of p-type SnSe as a function of temperature and net free-carrier concentration.
Figure~\ref{fig:S_ptype} shows the Seebeck coefficient as a function of temperature and carrier concentration along the \emph{a}, \emph{b}, and \emph{c} axes.
In the 300--600 K range the Seebeck coefficient data for the \emph{Pnma}-SnSe phase are almost isotropic and increase with temperature.
In this temperature range the Seebeck coefficients decrease at higher carrier concentrations
due to the reduction of the asymmetry of the density of states around the Fermi level for higher doping levels.
As temperature increases above 600 K, bipolar transport occurs and reduces the Seebeck coefficients.
The bipolar transport sets in at lower temperatures for lower carrier concentrations.
The temperature that the bipolar transport begins increases from 550 K to 700 K as the carrier concentration increases from 10$^{17}\text{ cm}^{-3}$ to 10$^{18}\text{ cm}^{-3}$.
For a net free-carrier concentration of 10$^{17}\text{ cm}^{-3}$ the Seebeck coefficient along \emph{a} and \emph{c} axes changes sign as temperature increases,
which implies that the thermally excited electrons start to dominate transport and the character of semiconductor changes from p-type to n-type.
For temperatures above 813 K, SnSe transitions to the \emph{Cmcm} phase and the calculated Seebeck coefficients are smaller than the \emph{Pnma} phase.
This is explained by the more important role of bipolar transport in the high-temperature \emph{Cmcm}
phase since it has a smaller band gap than the \emph{Pnma} phase.

\begin{figure}
\includegraphics[width=\columnwidth]{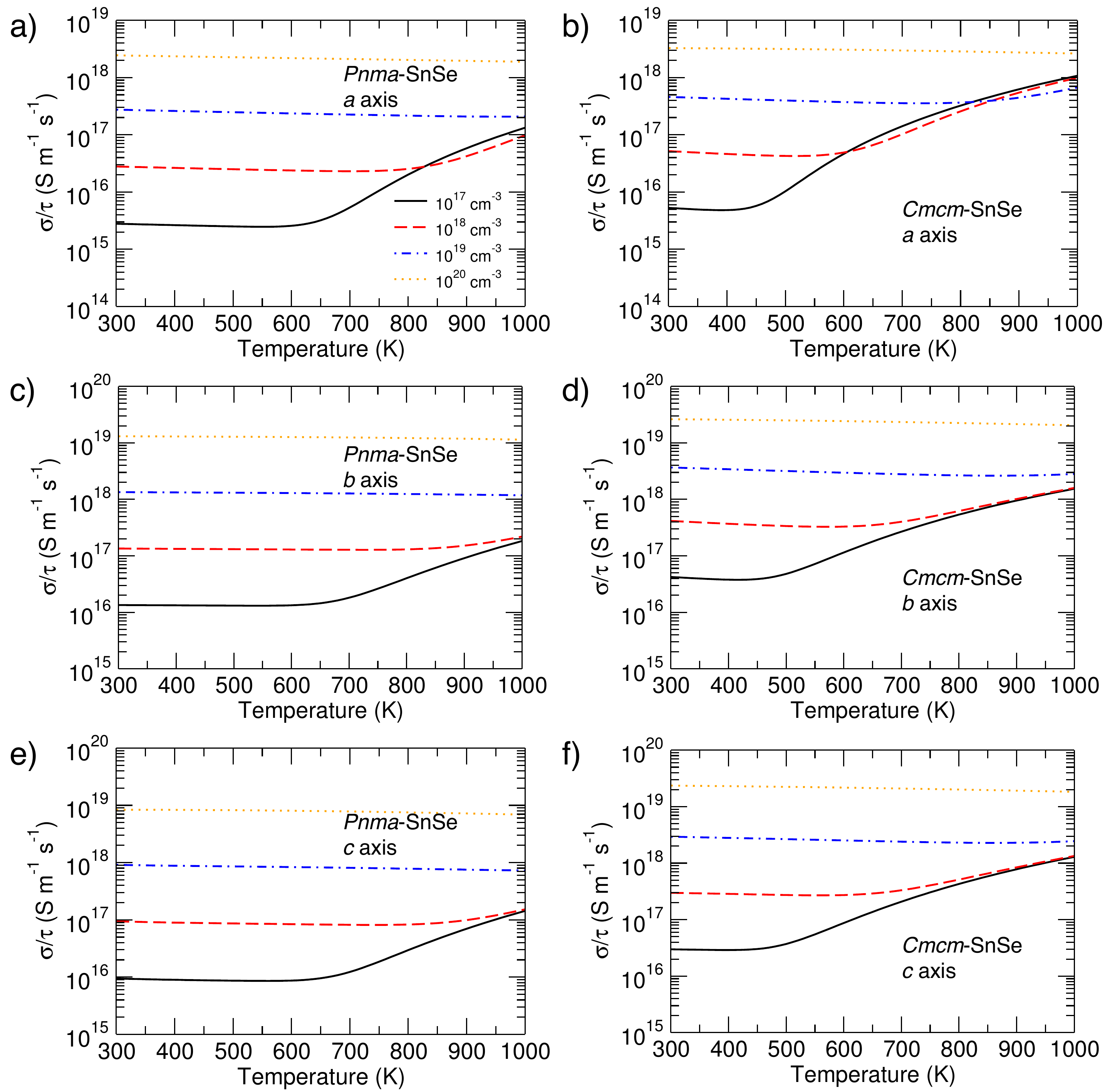}
\caption{
\label{fig:Sigma_ptype}
Electrical conductivity divided by the scattering time ($\sigma/\tau$) as a function of temperature and carrier concentration for the low-temperature (\emph{Pnma}) (a, c, e) and high-temperature (\emph{Cmcm}) phases (b, d, f) of SnSe along the \emph{a}, \emph{b}, and \emph{c} axes of the crystal structure.
}
\end{figure}

The calculated electrical conductivity of p-type SnSe divided by the scattering time ($\sigma/\tau$)
is plotted in Fig.~\ref{fig:Sigma_ptype}
as a function of crystal phase, crystal direction, net free-carrier concentration, and temperature.
The electrical conductivity is similar along the \emph{b} and \emph{c} axes but it is much smaller along the \emph{a} axis.
The thermally excited carriers dominate transport
and the electrical conductivity increases exponentially with temperature
for temperatures above the onset of bipolar transport.
The temperature onset of bipolar transport increases as the doping concentration increases because a larger number
of thermally excited carriers is needed to overcome the contribution by the doping carrier concentration to the electrical conductivity.
For temperatures above the phase transition (i.e., greater than 813 K) the \emph{Cmcm} phase is found to have much larger electrical conductivity
than the \emph{Pnma} phase because of the more important role of bipolar transport for \emph{Cmcm}-SnSe.

\begin{figure}
\includegraphics[width=\columnwidth]{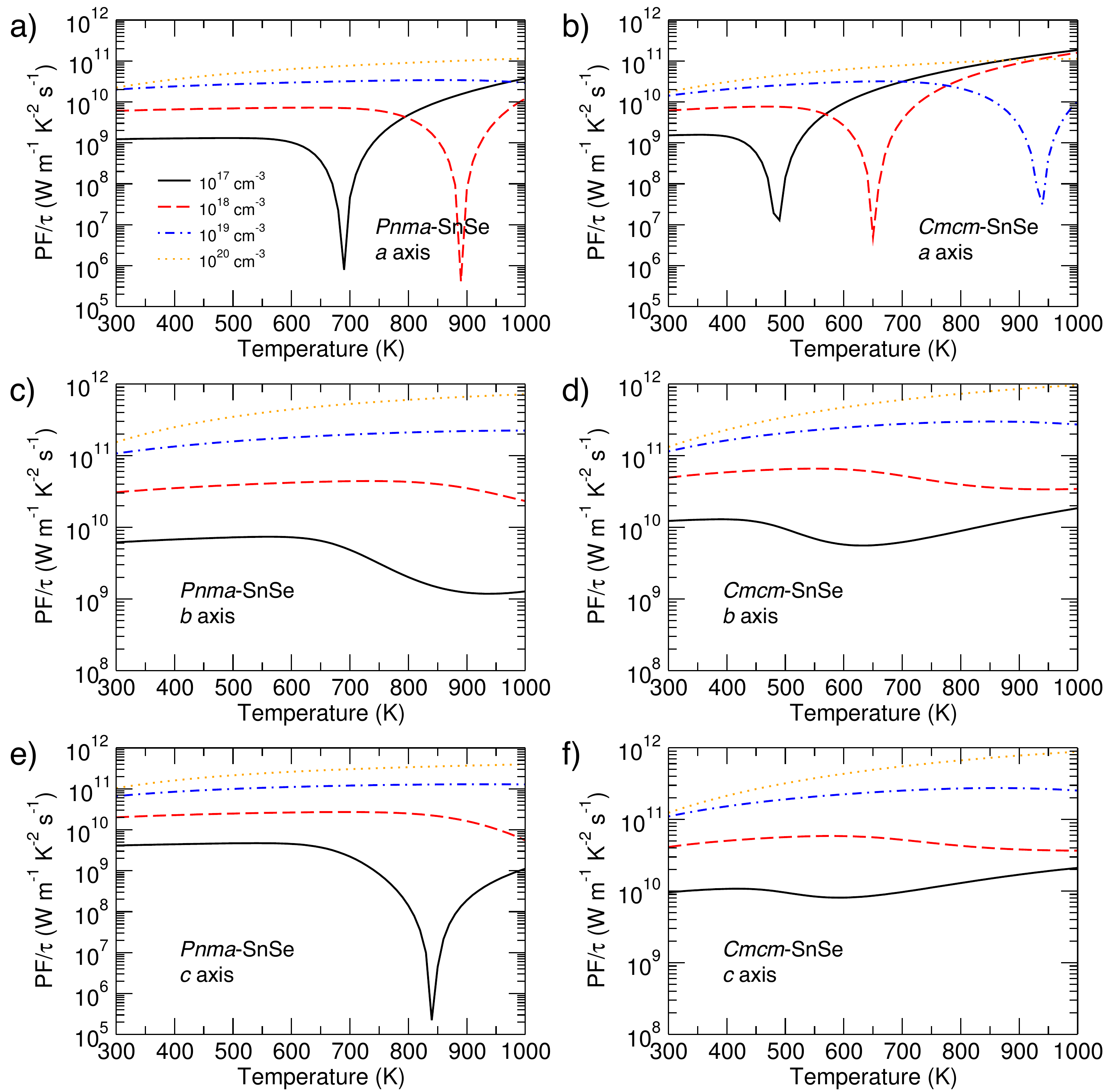}
\caption{
\label{fig:PF_ptype}
Power factor (PF) divided by the scattering time as a function of temperature and carrier concentration for the low-temperature (\emph{Pnma}) (a, c, e) and high-temperature (\emph{Cmcm}) phases (b, d, f) of SnSe along the \emph{a}, \emph{b}, and \emph{c} axes of the crystal structure.
}
\end{figure}

The power factor of \emph{Pnma} and \emph{Cmcm} SnSe divided by the scattering time is evaluated according to $\text{PF}/\tau = S^2 \sigma / \tau$
and is shown as a function of crystal direction, temperature, and net free-carrier density in Fig.~\ref{fig:PF_ptype}.
The power factor shows dips along the $a$ and $c$ axes because the Seebeck coefficient changes sign from positive to negative
and takes a zero value at the dip position.
The highest values are observed along the \emph{b} axis,
while the values are slightly smaller along the \emph{c} axis and much smaller along the \emph{a} axis.
This trend [$\text{PF}(b)>\text{PF}(c)>\text{PF}(a)$] agrees with the experimental observations.\cite{zhao2014ultralowthermal}
For high doping concentratons the power factor increases with increasing temperature.
The highest power factor is found for the highest carrier concentration (10$^{20}$ cm$^{-3}$) at the highest temperature (1000 K).

\begin{figure}
\includegraphics[width=\columnwidth]{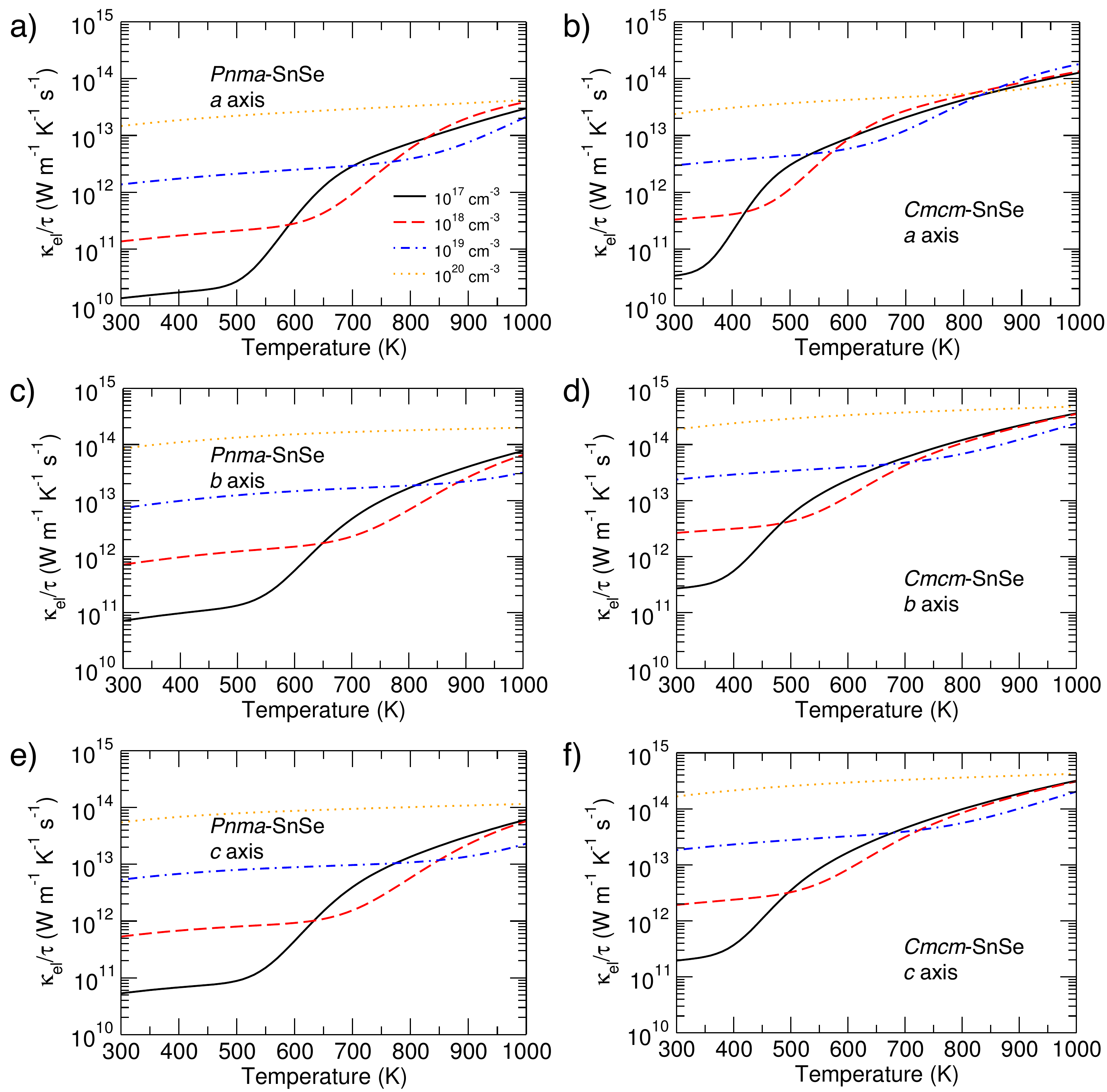}
\caption{
\label{fig:Kappae_ptype}
Electronic component of the thermal conductivity divided by the scattering time as a function for the low-temperature (\emph{Pnma}) (a, c, e) and high-temperature (\emph{Cmcm}) phases (b, d, f) of SnSe along the \emph{a}, \emph{b}, and \emph{c} axes of the crystal structure.
}
\end{figure}

Figure~\ref{fig:Kappae_ptype} shows the electron contribution to the thermal conductivity 
divided by scattering time
($\kappa_\text{el}/\tau$) for the two phases and the three crystal directions as a function of net free-carrier density and temperature.
The electron thermal conductivity increases with increasing temperature and doping concentration.
It takes higher values along the in-plane  $b$ and $c$ directions than the perpendicular $a$ direction, 
and it shows bipolar transport behavior similar to the electrical conductivity. 

\begin{figure}
\includegraphics[width=\columnwidth]{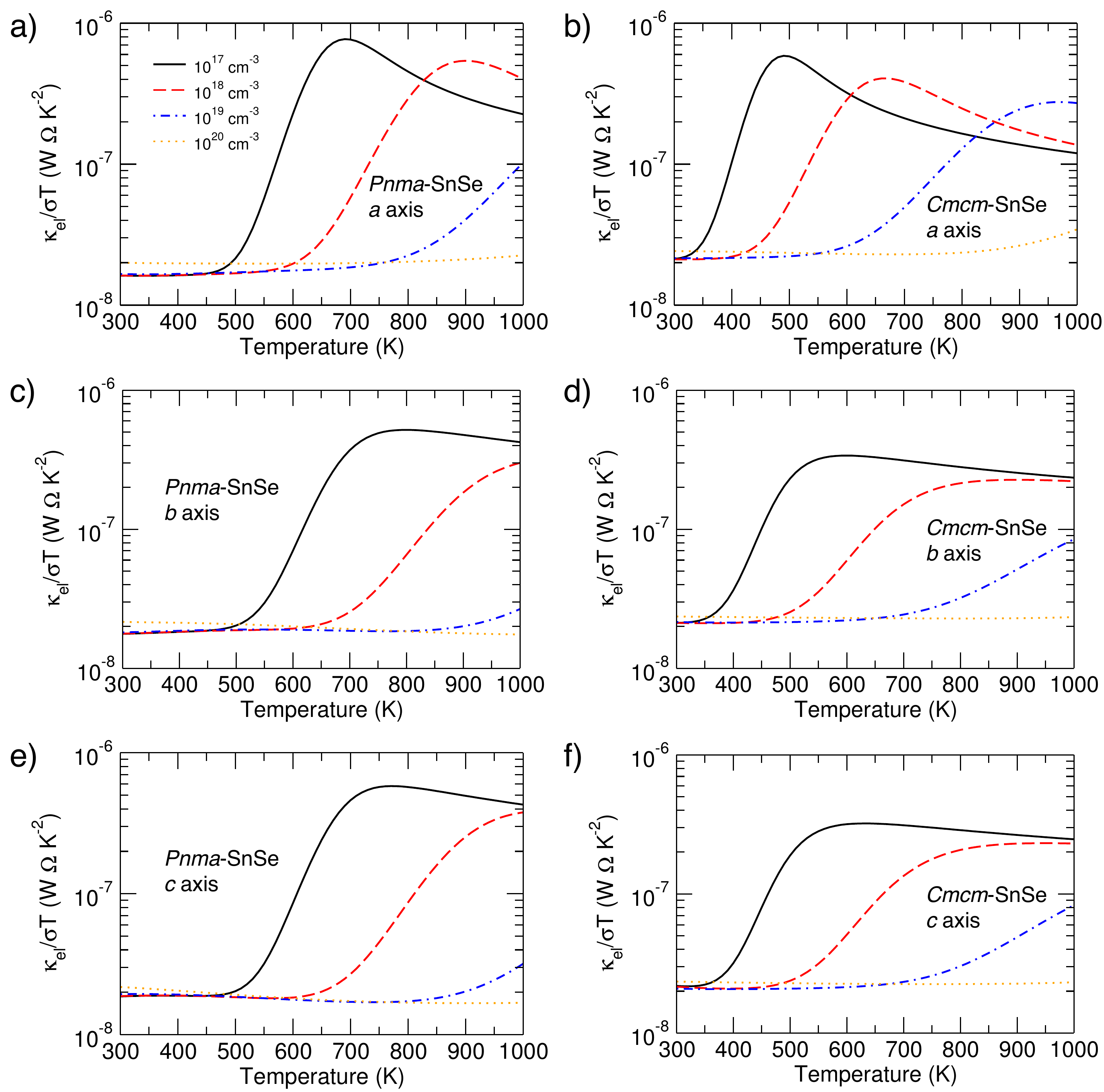}
\caption{
\label{fig:L_ptype}
The ratio of $\kappa_\text{el}$ to $\sigma T$ as a function of temperature and carrier concentration for the low-temperature (\emph{Pnma}) (a, c, e) and high-temperature (\emph{Cmcm}) phases (b, d, f) of SnSe along the \emph{a}, \emph{b}, and \emph{c} axes of the crystal structure.
}
\end{figure}

Figure~\ref{fig:L_ptype} shows the ratio of the thermal conductivity to the electrical conductivity multiplied by the temperature, $\kappa_\text{el}/\sigma T$,
as a function of direction, doping concentration, and temperature.
The values for this ratio in the limit of low temperature and low doping concentration are
1.7$\times$10$^{-8}$ W $\Omega$ K$^{-2}$,
1.8$\times$10$^{-8}$ W $\Omega$ K$^{-2}$,
and 1.9$\times$10$^{-8}$ W $\Omega$ K$^{-2}$ for the \emph{Pnma} phase along the \emph{a}, \emph{b}, and \emph{c} axes, respectively,
and 2.2$\times$10$^{-8}$ W $\Omega$ K$^{-2}$ for the \emph{Cmcm} phase along all three directions.
These are typical values for the Lorentz number of semiconductors for nondegenerate carriers.
As the temperature increases, bipolar transport increases the electronic thermal conductivity more 
than the electrical conductivity
due to the bipolar diffusion term
that is proportional to $\sigma_{\text{e}}\sigma_{\text{h}}/(\sigma_{\text{e}}+\sigma_{\text{h}})$, where $\sigma_{\text{e}}$ and $\sigma_{\text{h}}$
are the contributions to electrical conductivity by electrons and holes, respectively\cite{YangKappa}
and increases the value of this ratio.
This bipolar diffusion becomes significant if electrons and holes have large and similar electrical conductivities. 
As shown in Fig.~\ref{fig:L_ptype}, the $\kappa_\text{el}/\sigma{T}$ ratio increases by as much as a factor of 47
along the $a$ direction of the \emph{Pnma} phase at high temperature for a doping concentration of 10$^{17}$ cm$^{-3}$. A similar enhancement of the Lorenz number under bipolar transport has also been observed for bismuth telluride.\cite{0370-1301-69-2-310}
A constant value in the (1--2.4)$\times$10$^{-8}$ W $\Omega$ K$^{-2}$ for the Lorenz number is frequently used exprimentally to 
estimate the lattice and electronic contributions to the thermal conductivity.\cite{zhao2014ultralowthermal}
Our findings show that this assumption needs to be reexamined for SnSe if bipolar transport
affects the values of the electrical and electronic thermal conductivity at high temperatures.

\begin{figure}
\includegraphics[width=\columnwidth]{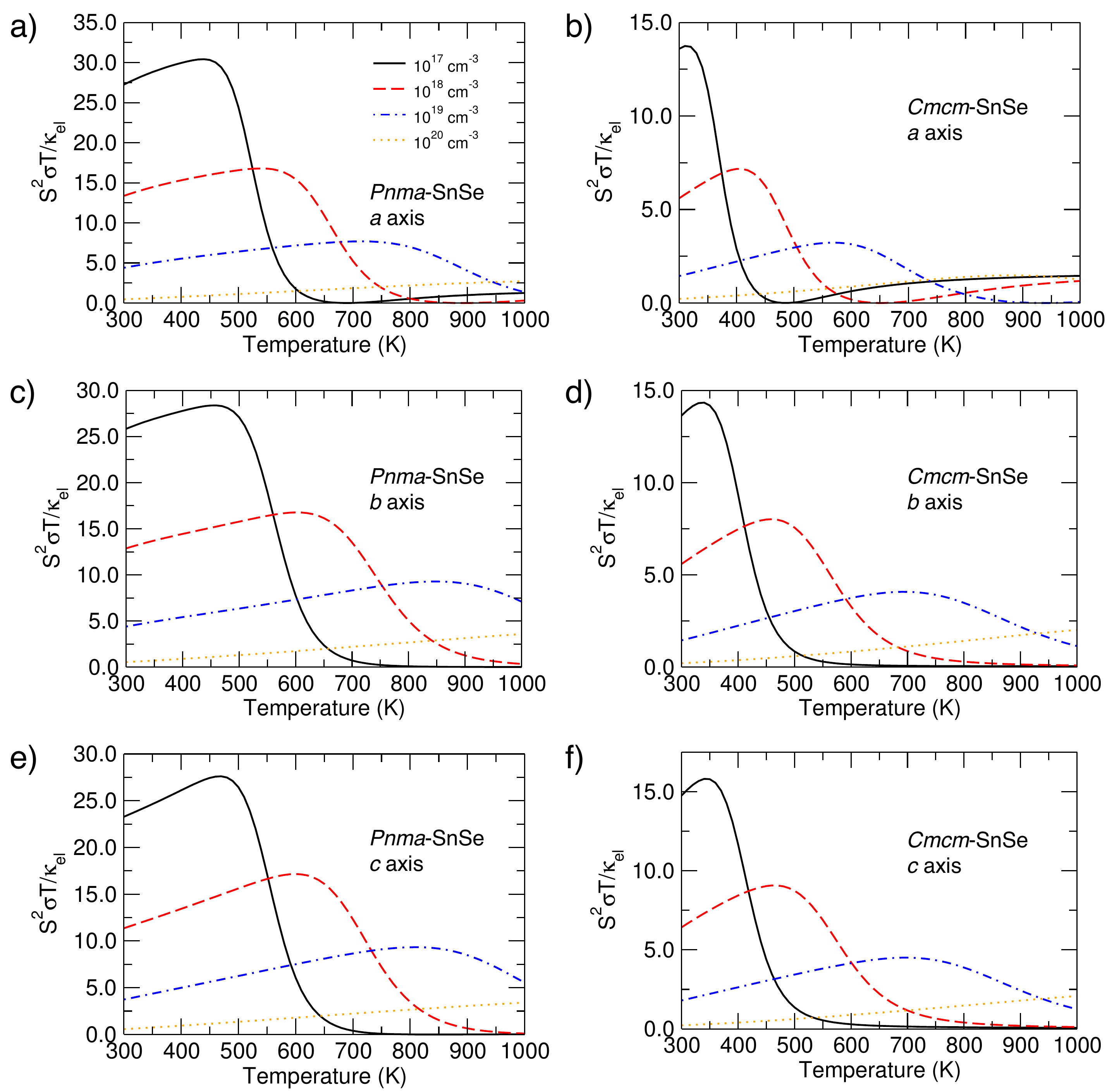}
\caption{
\label{fig:ZeT_ptype}
$S^2\sigma T/\kappa_\text{el}$ as a function of temperature and carrier concentration for the low-temperature (\emph{Pnma}) (a, c, e) and high-temperature (\emph{Cmcm}) phases (b, d, f) of SnSe along the \emph{a}, \emph{b}, and \emph{c} axes of the crystal structure.
}
\end{figure}

Figure~\ref{fig:ZeT_ptype} shows the ratio of $S^2\sigma{T}$ to $\kappa_\text{el}$ for \emph{Pnma}-SnSe and \emph{Cmcm}-SnSe as a function of crystal direction,
net free-carrier density, and temperature. This quantity is related to the figure of merit $ZT$ according to
\begin{align}\label{eq:ZT_equation}
ZT=\frac{S^2\sigma{T}}{\kappa_\text{el}}\frac{\kappa_\text{el}}{\kappa_\text{el}+\kappa_\text{L}}.
\end{align}
The quantity $S^2\sigma{T}/\kappa_\text{el}$ is independent of the constant scattering time and
is an upper limit to the figure of merit.
It approaches $ZT$
if the lattice contribution to the thermal conductivity is negligible compared to the electronic term.
For low doping concentration and low temperature,
the thermal conductivity is dominated by the lattice term and evaluating $ZT$
requires knowledge of the lattice thermal conductivity and the electron scattering time.
However, 
at high carrier concentration and high temperature the ratio $S^2\sigma{T}/\kappa_\text{el}$
is a good estimate of $ZT$
because the lattice thermal conductivity decreases with tempearture
and takes a remarkably low at a temperature above 700 K (0.20 W m$^{-1}$ K$^{-1}$)\cite{zhao2014ultralowthermal},
while the electronic thermal conductivity increases with 
increasing temperature (due to bipolar transport)
and increasing doping concentration.
Our results for \emph{Cmcm}-SnSe show that for net free-carrier concentration on the order of 10$^{17}$--10$^{18}$ cm$^{-3}$,
which is in the range of the experimental Hall measurements
of Zhao \emph{et al.},\cite{zhao2014ultralowthermal} the upper limit to $ZT$ along the $b$ and $c$ axes is much smaller than the
remarkably high $ZT$ values (as much as 2.6) reported experimentally.
This is an indication that the concentration of dopants in the experimental work of Zhao \emph{et al.}\cite{zhao2014ultralowthermal}
is larger than the Hall coefficients measured at room temperature (6$\times$10$^{17}$ cm$^{-3}$)
as the temperature increases beyond the phase transition.
For net free-hole concentrations in the 10$^{19}$--10$^{20}$ cm$^{-3}$ range  and temperatures above 700 K
the upper limit to $ZT$ takes values substantially larger than 1 along the $b$ and $c$ axes,
both for the \emph{Pnma} and the \emph{Cmcm} phase of SnSe.
It is desirable to dope SnSe with acceptors in the range of 10$^{19}$--10$^{20}$ cm$^{-3}$ to optimize the figure of merit at high temperature.

\section{Conclusions}

We investigated the band structure and electronic transport properties of both the low-temperature \emph{Pnma} and the high-temperature \emph{Cmcm} phase of SnSe.
We calculated the band gaps and carrier effective masses and we found that both phases exhibit multiple local band extrema near the band edges that
need to be considered when evaluating the thermoelectric properties for this material.
Our calculated transport coefficients shed light into recent experimental measurements that reported a remarkably high figure-of-merit value (2.6) for \emph{Cmcm}-SnSe.
Our results predict that SnSe would show optical thermoelectric performance at high temperature
when doped in the 10$^{19}$--10$^{20}$ cm$^{-3}$ range. 

\acknowledgments
We thank C. Uher, H. Chi, and P. F. P. Poudeu for useful discussions.
This work was supported as part of CSTEC, an Energy Frontier Research Center funded by the U.S. Department of Energy, Office of Science, Basic Energy Sciences under Award \# DE-SC0000957.
Computational resources were provided by the National Energy Research Scientific Computing Center, which is supported by the Office of Science of the U.S. Department of Energy under Contract No. DE-AC02-05CH11231.

%
\end{document}